\documentclass[useAMS,usenatbib]{mn2e}

\usepackage{epsfig}

\begin{document}

\title{Thermal instability with the effect of cosmic-ray diffusion}

\author[M. Shadmehri]{Mohsen Shadmehri\thanks{E-mail:
mshadmehri@thphys.nuim.ie (MS); }\\
Department of Mathematical Physics, National University Ireland, Co Kildare, Maynooth, Ireland}

\maketitle

\date{Received ______________ / Accepted _________________ }

\begin{abstract}
We study dynamical effects of cosmics rays (CRs) on the thermal instability in the linear regime. CRs and the thermal plasma are treated as two different interacting fluids, in which CRs can diffuse along the magnetic field lines. We show that growth rate of the magnetothermal condensation mode is reduced because of the existence of CRs and this stabilizing effect depends on the diffusion coefficient and the ratio of the CRs pressure to gas pressure. Thus, a slower rate of structure formation via thermal instability is predicted when CRs are considered.

\end{abstract}

\begin{keywords}
ISM: cosmic rays - instabilities - stars: formation
\end{keywords}
\section{Introduction}
\label{sec:1}

Birth of the stars and the complex observable patterns within the interstellar medium (ISM) are mainly understood by the instabilities that may start with small amplitude perturbations leading to highly nonlinear processes, i.e. turbulence \citep[e.g.,][]{scalo2004, maclow2004}. In fact, among the most important physical ingredients in any theory of star formation one can mention self-gravity, magnetic field and net cooling function. However, there is another significant physical factor that its {\it dynamical} role has not been considered in most scenarios of star formation until rather recently: Cosmic Rays (CRs).

Equipartition exists between energy densities of thermal gas, magnetic field, turbulence and CRs \citep[e.g.,][]{gaisser90, ferriere2001}. The energy equipartition encourages one to expect significant dynamical effects of CRs on the structure formation in ISM as has been realized by Parker in a simplified model \citep{parker66}. It is believed that giant cloud complexes in the spiral arms of galaxies are forming via Parker instability \citep[e.g.,][]{mous74, elmegreen86}. More recent studies show that the growth rate of the Parker instability is very sensitive to the CR pressure and the diffusion coefficient \citep[e.g.,][]{ptuskin, hanasz97, hanasz2000, kuwabara2004, kuwabara2006}. CRs have also a vital role in propagation of MHD waves. \citet{Lo2007} studied the the stability of a cosmic-ray plasma system. This stability analysis has been generalized to a four-fluid cosmic-ray-MHD system which comprises magnetized thermal plasma, cosmic rays, forward and backward propagating Alfvén waves \citep{ko2009}.

Interactions of CRs may also operate as a global heating mechanism in ISM \citep*[e.g.,][]{field69, goldsmith69}. Recently, \citet*{zadeh2007} proposed that due to heating of the gas clouds in the central regions of our Galaxy by an enhanced flux of CR electrons, the rate of star formation in these star forming sites decreases and a similar mechanism may operate in the nuclei of the other galaxies. In another related study, \citet*{fatuzzo2006} studied the interplay between molecular clouds and the irradiation by CRs produced by supernova remnants. This increased CRs flux has important consequences for star formation. In particular, a higher ionization level due to CRs lead to a longer ambipolar diffusion time scale and therefore slower star formation rate \citep{fatuzzo2006}. However, none of these studies considered the possible  dynamical effects of CRs on the star forming regions.

A linear analysis of non-magnetized thermal gas and CR has revealed that CRs suppress the growth of small amplitude perturbations \citep{ptuskin, begelman94, wagner2005}. \citet{ptuskin} (hereafter KP83) studied stability of ISM consisting of regular and turbulent magnetic fields, thermal gas and CRs. Although stabilizing effect due to the existence of CRs has been shown,  heating-cooling of the system has not been considered by KP83 and a polytropic equation of state is used instead of the energy equation for the gas component. Moreover, their analysis resembles to the classical Parker instability, in which the initial unperturbed state is determined by the balance of the forces rather than thermal equilibrium states as we will consider in our study. Therefore, analysis of \citet{ptuskin} can not address stability of a system with CRs when the net cooling function is playing a significant role like structure formation in HI regions due to the thermal condensation \citep[e.g.,][]{fukue}. In another study, \citet{wagner2005} derived stability condition of a non-magnetized thermal system for a few simplified cases. However, growth rate of the perturbations has not been studied by \citet{wagner2005}. In this study, we follow a similar approach but in a magnetized case considering the net cooling function of the system and the dynamical effects of CRs. In this regard, our study is different from KP83 who did not take into account thermal behavior of the system and as we will show growth rate of the perturbations and the criteria of stability are completely different in comparison to KP83. Also, our study is  complementary to \citet{wagner2005} who did not consider magnetic field and the growth rate of the perturbations has not been calculated in detail. It is assumed that CRs particles can diffuse along the magnetic field lines \citep[see also,][]{Lerche}. In the next section basic equations of a magnetothermal system including CRs are presented. A dispersion relation is obtained for the linear perturbations in section \ref{sec:LP}. Analysis of the the growth rate of the condensation modes is done in section \ref{sec:A}. We conclude by a summary of the results and possible implications in the final section.

\section{General Formulation}
\label{sec:GF}
There are three different approaches to study the dynamics of CRs. In the particle-particle approach, the plasma and CRs  are considered as particles that may interact with each other via complicated processes. In a simpler approach, known as fluid-particle, the plasma is treated as a fluid, though CRs are still described as particles. The simplest approach is the fluid-fluid approach in which CRs and the thermal gas are described by different interacting fluids. The hydrodynamic  approach can not provide the spectrum of CRs, however, it is a good approximation to use to analyze dynamics of a plasma with CRs \citep[e.g.,][]{drury81,drury83}. We adopt the hydrodynamic approach to study effects of CRs on the thermal instability.

We also follow the same steps as in \citet*{field65}, except  that CRs are considered as a fluid, and diffusion is considered only along magnetic field lines. For simplicity we neglect ionization and heating by CRs, since their effects in the absence of the dynamical role of CRs are well understood \citep[e.g.,][]{goldsmith69}.

The basic equations are
\begin{equation}\label{eq:cont}
\frac{d\rho}{dt}=-\rho \nabla.{\bf v},
\end{equation}
\begin{equation}\label{eq:motion}
\rho \frac{d{\bf v}}{dt}=\frac{1}{4\pi} (\nabla \times {\bf B}) \times {\bf B} - \nabla (p+p_{\rm cr}),
\end{equation}
\begin{equation}
\frac{\partial {\bf B}}{\partial t}=\nabla \times ({\bf v} \times {\bf B}),
\end{equation}
\begin{equation}
\nabla . {\bf B}=0,
\end{equation}
\begin{equation}
\frac{1}{\gamma -1}\frac{dp}{dt}-\frac{\gamma}{\gamma -1}\frac{p}{\rho}\frac{d\rho}{dt}+\rho \Omega - \nabla . [K_{\parallel}\nabla_{\parallel} T + K_{\perp} \nabla_{\perp} T] =0,
\end{equation}
and
\begin{equation}\label{eq:CRS}
\frac{1}{\gamma_{\rm cr}-1}\frac{dp_{\rm cr}}{dt}-\frac{\gamma_{\rm cr}}{\gamma_{\rm cr}-1} \frac{p_{\rm cr}}{\rho} \frac{d\rho}{dt}+\nabla . {\bf \Gamma} =0,
\end{equation}
where $d/dt = \partial / \partial t + {\bf v}.\nabla$ is the Lagrangian time derivative,
\begin{equation}
{\bf \Gamma}=-\kappa_{\parallel} {\bf b} ({\bf b}.\nabla p_{\rm cr}),
\end{equation}
is the diffusive flux of cosmic-ray energy and $\kappa_{\parallel}$ is diffusion coefficient along magnetic field lines. All the variables have their usual meaning. Also, $\bf b$ is a unit
 vector along the magnetic field lines, i.e. ${\bf b}= {\bf B}/B$. The adiabatic indices of the thermal gas and cosmic rays are denoted by $\gamma$ and $\gamma_{\rm cr}$, respectively. Also, $\Omega$ represents the
 energy losses minus energy gains per unit mass. The coefficient of thermal conductivity $K$ has the values $K_{\parallel}$ and $K_{\perp}$ in
 directions parallel to and perpendicular to the magnetic field $\bf B$. Finally, we can write equation of state as
\begin{equation}\label{eq:STATE}
p=\frac{R}{\mu} \rho T,
\end{equation}
where $R$ is the gas constant and $\mu$ represents the molecular weight.

The energy equation of CRs in its complete form includes also an extra term corresponding to an effective CRs energy loss (see equation (13) in \citet*{Lerche}). But this term is neglected in our analysis simply based on a time scale argument. We neglect the
ionization energy losses of the cosmic rays because  these are surely on a
much longer time scale than we are interested.  In fact, CR pressure is mainly from mildly relativistic protons, and that these
have very low ionization energy loss rates and also low nuclear
collision rates. Therefore, typically the ionization and nuclear loss
time scales are around $10^8$ or more years for normal interstellar
values.  But typical cooling time scale is around $10^6$ years. Also, there is an effect due to second-order Fermi acceleration of cosmic rays ( see equations (15) and (13) in \citet*{Lerche}), which in fact would turn the cosmic ray loss time into a cosmic ray gain time. This effect is not so easy to discard, as this acceleration is caused by the same MHD waves that control the spatial diffusion of cosmic rays. In fact it is known \citep{schi} that the product of spatial diffusion time $\tau_{\rm D}$ and the acceleration time scale $\tau_{\rm F}$ equals a constant that depends  on the spectral index of the turbulence spectrum and
$(L/v_{\rm A})^{2}$, where $L$ denotes the size of the system and $v_{\rm A}$ is the Alfven speed (see equation (86) in \citet*{schi}). The time scales are related approximately $\tau_{\rm D} \tau_{\rm F} \cong (L/v_{\rm A})^{2}$ \citep{schi}, where the spatial diffusion time scale is $\tau_{\rm D} = L^{2}/\kappa$. For a typical cooling ISM we can assume $L=10^{20}$ cm, $B_{0}=10^{-6}$ G and the ion density is $n_{i}=10^{-2}$ cm$^{-3}$. So, we obtain $\tau_{\rm D} = 10^{12}$ s and $\tau_{\rm F} = 7.2\times 10^{17}$ s and so, $\tau_{\rm F} \gg \tau_{\rm D}$ which means we are indeed in a parameter range to neglect the term corresponding to an effective CRs energy loss. Thus, it seems our approach is fine. We actually have no source
term either, but we are interested in the effect of a pre-existing and
stable CR population on the magnetothermal condensation modes.  To introduce additional time
scales related to sources and sinks is a nonsense unless they are
comparable to the time scales we are studying.

Note that in the hydrodynamic approach of CR, the energy spectrum of the particles of CR is not considered in detail. Moreover, type of the CR particles is not specified in this analysis and they can consist of electrons or protons in any combination. In comparison to the classical magnetothermal instability analysis \citep{field65}, there is an extra pressure term due to the CRs in the equation of motion (\ref{eq:motion}). So, one may expect an stabilization effect of CRs. But equation (\ref{eq:CRS}) describes advection of CRs by flowing gas and the diffusion of CRs along the magnetic field lines through the thermal gas.  Little is known about the diffusion coefficient $\kappa_{\parallel}$ and its possible dependence on the physical variables of the system. As explained in \citet*{Lerche} the parallel spatial cosmic ray diffusion coefficient
$\kappa_{\parallel }$ in a cosmic ray fluid theory is an effective diffusion coefficient averaged over
all cosmic ray momenta. Depending on the actual momentum distribution function of cosmic rays and
the momentum dependent cosmic ray diffusion coefficient, its value can be very different from
the standard value $10^{28}$ cm$^2$ s$^{-1}$ \citep[see also,][]{berezinski90, ptuskin2001}.

\section{Linear perturbations}
\label{sec:LP}

We completely neglecting gradients in the background medium and the velocity is zero in equilibrium state $\rho_0$, $T_0$, $p_0$, $p_{\rm cr0}$. Also, we assume $\Omega(\rho_0, T_0)=0$. By
perturbing  of the form
$
\delta X (\textbf{r},t) = \delta \bar{X} \exp
 (\omega t + i \textbf{k} \cdot \textbf{r}),
$
the equations (\ref{eq:cont})-(\ref{eq:STATE}) become
\begin{equation}
\omega \delta \bar{\rho} + i \rho_{0} {\bf k}.\delta {\bar{\bf{v}}}=0,
\end{equation}
\begin{equation}
\omega \rho_{0} \delta\bar{{\bf v}} + i {\bf k} \delta\bar{p} + i {\bf k} \delta\bar{p}_{\rm cr} + i ({\bf B}_{0}.\delta \bar{{\bf B}})\frac{{\bf k}}{4\pi} - i ({\bf k}.{\bf B}_{0}) \frac{\delta\bar{\bf {B}}}{4\pi}=0,
\end{equation}
\begin{equation}
\omega \delta\bar{{\bf B}} + i {\bf B}_{0} ({\bf k}.{\delta\bar {\bf{v} }})- i ({\bf k}.{\bf B}_{0}) \delta\bar{{\bf v}}=0,
\end{equation}
\begin{displaymath}
\frac{\omega}{\gamma -1} \delta \bar{p} - \frac{\omega\gamma p_{0}}{(\gamma -1)\rho_{0}} \delta\bar{\rho} + \rho_{0} \Omega_{\rho} \delta\bar{\rho} + \rho_{0} \Omega_{T} \delta\bar{T}
\end{displaymath}
\begin{equation}
+(K_{\parallel} k_{\parallel}^{2} + K_{\perp} k_{\perp}^{2}) \delta\bar{T}=0,
\end{equation}
\begin{equation}
\frac{\omega}{\gamma_{\rm cr} -1} \delta \bar{p}_{\rm cr} - \frac{\omega\gamma_{\rm cr} p_{\rm cr0}}{(\gamma_{\rm cr} -1)\rho_{0}} \delta\bar{\rho} + \kappa_{\parallel} \frac{({\bf k}.{\bf B}_{0})^2}{B_{0}^{2}} \delta \bar{p}_{\rm cr}=0,
\end{equation}
\begin{equation}
\frac{\delta \bar{p}}{p_0} - \frac{\delta \bar{\rho}}{\rho_0}-\frac{\delta \bar{T}}{T_0}=0
\end{equation}
Note that the derivative $\Omega_{\rho}=(\partial\Omega/\partial\rho)_{T}$ and $\Omega_{T}=(\partial\Omega/\partial T)_{\rho}$ are evaluated for the equilibrium state.

Then, we introduce the coordinate system $\textbf{\textit{e}}_x$,
$\textbf{\textit{e}}_y$, and $\textbf{\textit{e}}_z$ specified by
\begin{equation}\label{coord}
\textbf{\textit{e}}_z=\frac{\textbf{B}_0}{B_0}\quad,\quad\textbf{\textit{e}}_y=\frac{\textbf{B}_0\times\textbf{k}}
{|\textbf{B}_0\times\textbf{k}|}\quad,\quad\textbf{\textit{e}}_x=\textbf{\textit{e}}_y\times\textbf{\textit{e}}_z.
\end{equation}
Also, we introduce the following wavenumbers
\begin{displaymath}
k_{\rho}=\mu (\gamma -1) \rho_{0}\Omega_{\rho} (Rc_{\rm s} T_{0})^{-1}, k_{T} = \mu (\gamma -1 ) \Omega_{T} (Rc_{\rm s})^{-1},
\end{displaymath}
\begin{displaymath}
k_{K_{\parallel}} = [\mu (\gamma -1) K_{\parallel}]^{-1} (Rc_{\rm s}\rho_{0}),
\end{displaymath}
\begin{equation}
k_{K_{\perp}} = [\mu (\gamma -1) K_{\perp}]^{-1} (Rc_{\rm s}\rho_{0}).
\end{equation}
Now, we can write the dispersion equation using the following non-dimensional quantities,
\begin{displaymath}
\Gamma = \frac{\omega}{k c_{\rm s}}, \sigma_{\rho} = \frac{k_{\rho}}{k}, \sigma_{T}=\frac{k_{T}}{k}, \sigma_{K_{\parallel}}=\frac{k}{k_{K_{\parallel}}}, \sigma_{K_{\perp}}=\frac{k}{k_{K_{\perp}}}.
\end{displaymath}

\begin{figure*}
\epsfig{figure=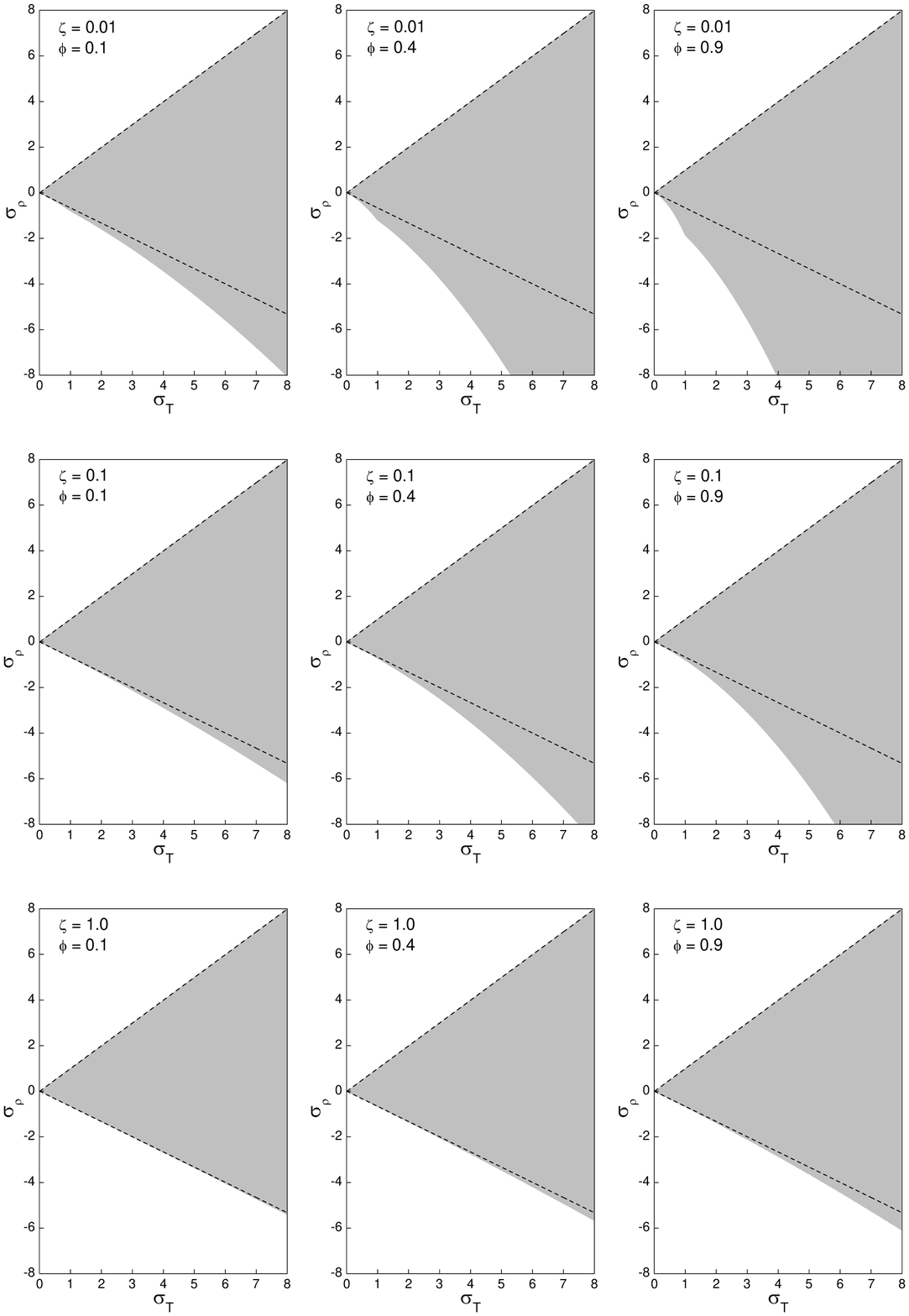,angle=0,scale=0.75}
\caption{Domains of stability are shown as light gray areas for $\zeta=0.01$ (top), $\zeta=0.1$ (middle) and $\zeta=1.0$ (bottom) when $\psi=100$, $\alpha=1$, $\gamma=5/3$ and $\gamma_{cr}=4/3$. The region between two dashed lines is corresponding to the stable domain without CRs.}
\label{fig:ff1}
\end{figure*}

Also, possible effects of CRs diffusion are appeared in our final dispersion relation through the non-dimensional parameters $\phi$ and $\psi$ as
\begin{equation}
\phi = \frac{p_{cr0}}{p_0},
\end{equation}
\begin{equation}
\psi=\psi_{0}\frac{k}{k_{\rho}}
\end{equation}
where $\psi_{0}=(\gamma_{cr} -1)k_{\rho}(c_{s}/\kappa_{\parallel})^{-1}$. Defining a new wavenumber $k_{c}=(\gamma_{cr} -1)^{-1} (c_{s}/\kappa_{\parallel})$, then we have $\psi_{0}=k_{\rho}/k_{c}$. Also, this parameter can be re-written as $\psi_{0}\approx \mu (\gamma_{cr}-1) \kappa_{\parallel} / c_{s}^{2} \tau_{cool}$, where $\tau_{cool}$ is the cooling time-scale. Obviously, the non-dimensional parameter $\psi_{0}$ shows the level of diffusion of CRs. When there is no diffusion and the thermal gas and CRs are well coupled together,  we have $\psi=0$. But as this parameter increases CRs are diffusing more along the magnetic field lines, the level of the coupling becomes weaker. In our analysis, these non-dimensional parameters are the key input parameters to explore possible effects of CRs on the thermal instability.

Therefore, the characteristic equation becomes
\begin{displaymath}
\Gamma^6+(\sigma_{T}+\sigma_{K}+\psi \zeta)\Gamma^5+ [\psi \zeta(\sigma_{T}+\sigma_{K})+1+\alpha
\end{displaymath}
\begin{displaymath}
+\gamma_{cr}\phi/\gamma]\Gamma^4 + [ (1+\gamma\alpha+\gamma_{cr}\phi)(\sigma_{T}+\sigma_{K})/\gamma- \sigma_{\rho}/\gamma
\end{displaymath}
\begin{displaymath}
 + \psi\zeta (1+\alpha)]\Gamma^3 + [ \psi \zeta (1+\gamma\alpha) (\sigma_{T}+\sigma_{K})/\gamma-\psi \zeta\sigma_{\rho}/\gamma
\end{displaymath}
\begin{displaymath}
+ \alpha \zeta (1+\gamma_{cr}\phi/\gamma) ] \Gamma^2 + [ \alpha\zeta (1+\gamma_{cr}\phi)(\sigma_{T}+\sigma_{K})/\gamma
\end{displaymath}
\begin{equation}\label{eq:disper}
-\alpha\zeta\sigma_{\rho}/\gamma + \alpha\psi \zeta^2]\Gamma + \alpha\psi \zeta^2 (\sigma_{T}+\sigma_{K}-\sigma_{\rho})/\gamma =0,
\end{equation}
and $\zeta = \cos^{2}\theta$ and $\theta$ is the angle between ${\bf B}_{0}$ and ${\bf k}$. Also, we have
$\sigma_{K}=\sigma_{K_{\parallel}} \zeta + \sigma_{K_{\perp}} (1-\zeta)$ and $\alpha=(v_{\rm A}/c_{\rm s})^{2}$, where $v_{\rm A}$ is the Alfven velocity.
%

\section{analysis}
\label{sec:A}

\begin{figure}
\vspace{-40pt}
\epsfig{figure=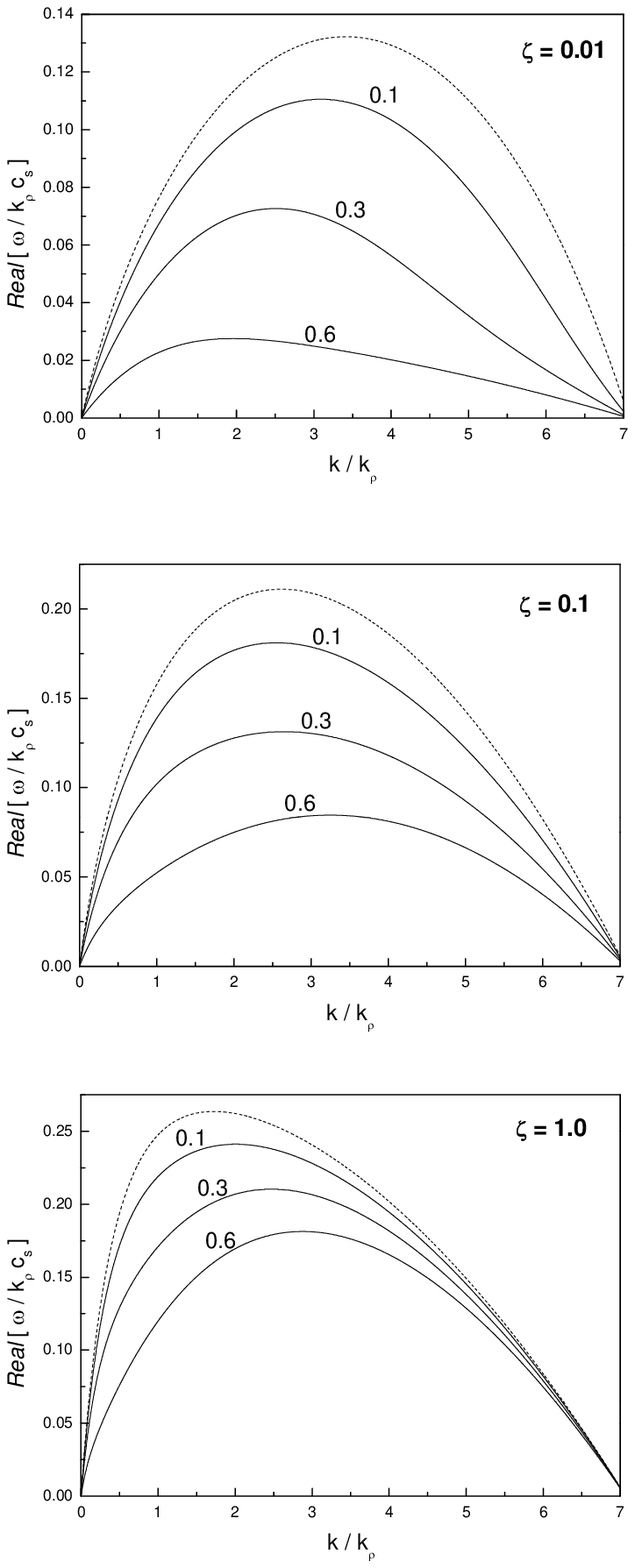,angle=0,scale=1.0}
\caption{Growth rate of thermal condensation mode versus wavenumber of the perturbations when $\alpha=1$, $\gamma=5/3$, $\gamma_{cr}=4/3$, $\psi_{0}=0.1$,  $\sigma_{T}/\sigma_{\rho} = 1/2$ and $\sigma_{\rho} \sigma_{K} = 0.01$. Dashed line represents growth rate in the case without CRs. Each curve is labeled by $\phi$.}
\label{fig:f1}
\end{figure}

\begin{figure}
\vspace{-40pt}
\epsfig{figure=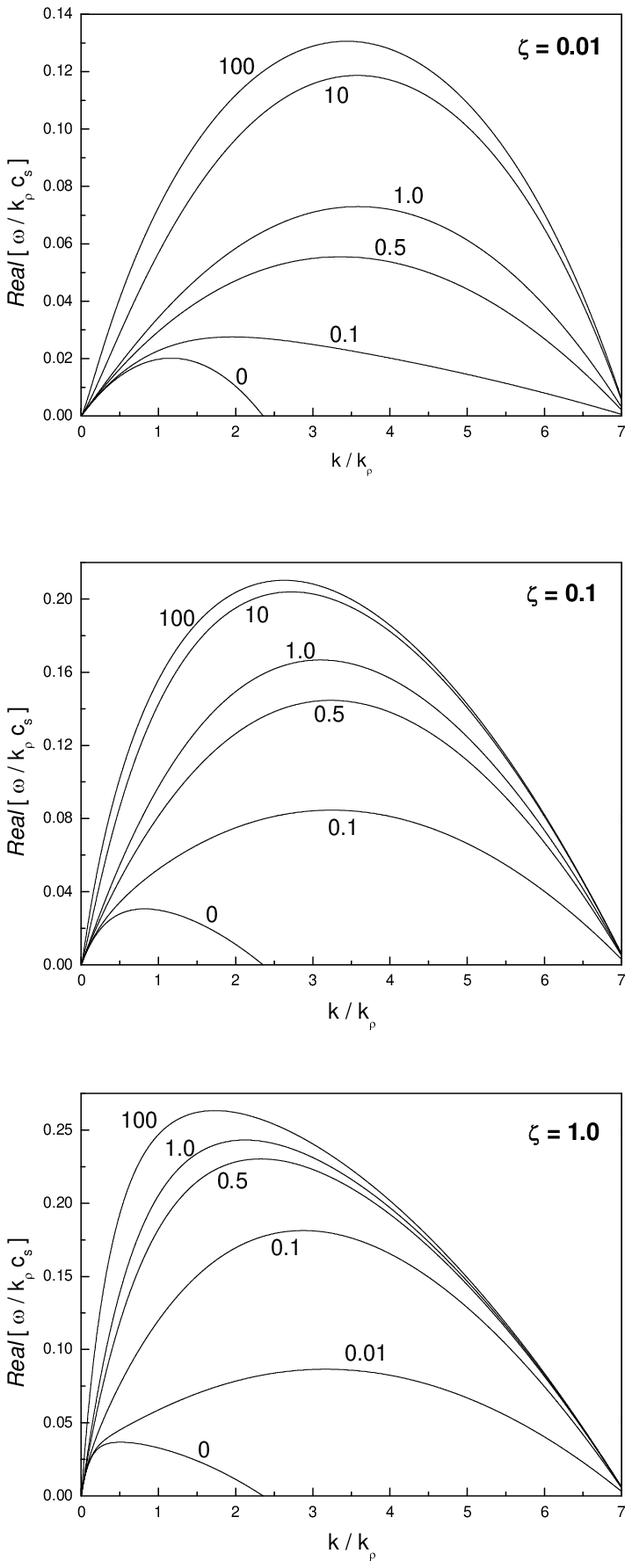,angle=0,scale=1.0}
\caption{Growth rate of thermal condensation mode versus wavenumber of the perturbations when $\alpha=1$, $\gamma=5/3$, $\gamma_{cr}=4/3$, $\phi=0.6$,  $\sigma_{T}/\sigma_{\rho} = 1/2$ and $\sigma_{\rho} \sigma_{K} = 0.01$. Each curve is labeled by $\psi_{0}$.}
\label{fig:f2}
\end{figure}
Equation (\ref{eq:disper}) describes magneto-thermal waves and unstable modes in a thermal system with CRs. If we set $\phi=\psi=0$, this algebraic equation reduces to the standard characteristic equation of magneto-thermal instability without CRs \citep[e.g.,][]{field65}. We are mainly interested in roots of equation (\ref{eq:disper}) that are real and positive (i.e., condensation modes). Qualitative analysis of the roots of the algebraic polynomial equations is generally done by Hurwitz analysis. In a case without magnetic field, \citet{wagner2005} derived conditions of the stability of the system when $\phi \gg 1$ or $\phi \ll 1$. However, we found it difficult to do such an analysis analytically and present the stability criteria in a simplified closed form (except for $\zeta=0$) because of the complicated coefficients of the characteristic equation (\ref{eq:disper}). But we study domains of instability numerically. In analogy to \citet{field65}, we show domains of the instability based on the characteristic equation (\ref{eq:disper}) in a system of coordinates defined by $\sigma_{\rho}$ and $\sigma_{T}$. Also, in order to study the influence of various parameters on the growth rate of the condensation modes, we solve the dispersion equation (\ref{eq:disper}) numerically to locate the roots of the equation against the wave number for several values of the parameters.

Domains of the stability are shown in Fig. \ref{fig:ff1}. For simplicity we consider cases without thermal conduction. In Fig. \ref{fig:ff1}, the vertical axis is related to the density derivative of the net cooling function, i.e. $\sigma_{\rho}$, and the horizontal axis is the normalized form of the temperature derivative of the cooling function, i.e. $\sigma_{T}$. The nondimensioanl  parameter of the diffusion of CRs is $\psi=100$ and the ratio $\phi$ varies as $0.1$, $0.4$ and $0.9$ and also we assume $\alpha=1$, $\gamma=5/3$ and $\gamma_{cr}=4/3$. Top, middle and bottom plots of Fig. \ref{fig:ff1} are corresponding to $\zeta=0.1$, $0.01$ and $1$, respectively. Light gray regions in this Figure are domains of the stability. In order to make easier comparison, the area between two dashed lines corresponds to the domain of the stability in a case without CRs. We see that domains of the stability extend due to the existence of CRs, in particular, for nearly transverse perturbations this extension is very significant. There are stable regions for nearly perpendicular perturbations, for which parallel perturbations are not stable. But in the absence of CRs, domains of stability  are the same for all the cases. Therefore, existence of CRs may lead to the formation of highly elongated or even filamentary  structures.

As we mentioned for transverse perturbations, i.e. $\zeta=0$, we can present a simple analytical criteria for stability of the system. In this case, the characteristic equation (\ref{eq:disper}) reduces to a third degree algebraic equation for which one can simply do Hurwitz analysis. Criteria of the stability is the following inequality
\begin{equation}\label{eq:stable}
\sigma_{T}+\sigma_{K} > \frac{\sigma_{\rho}}{1+\gamma_{cr} \phi + \gamma \alpha}.
\end{equation}
Obviously, in the absence of CRs and magnetic field the above condition reduces to the classical thermal stability criteria \citep{field65}.

Now, we study condensation modes numerically.  We take parameters $\alpha =1$, $\sigma_{T}/\sigma_{\rho} = 1/2$ and $\sigma_{\rho} \sigma_{K} = 0.01$ for easier comparison to \citet{field65}. Also, the other parameters are $\gamma=5/3$ and $\gamma_{cr}=4/3$. Effects of the initial ratio of the CRs pressure to the gas pressure on the growth rate of condensation modes is explored in Fig. \ref{fig:f1} by fixing $\psi_{0}=0.1$ and changing $\phi$ from $0.1$ to $0.6$. In the case of no CRs, the growth rate is represented by a dashed curve. Top plot is for perturbations that are nearly perpendicular to the initial magnetic field line, i.e. $\zeta=0.01$. But the bottom plot shows growth rate for the cases with  the perturbations along the initial magnetic field line, i.e. $\zeta=1$. Note that each curve is labelled by corresponding ratio $\phi$. Reduced growth rates are seen in Fig. \ref{fig:f1} due to the existence of CRs, irrespective of the direction of the perturbations. Also, reduction of the growth rate is larger as the ratio of the unperturbed CRs pressure to gas pressure increases. This stabilizing effect of CRs can be easily understood by noting to the fact that CRs pressure is acting as an extra support to help the gas pressure against the unstable perturbations in the momentum equation.

Comparing to a case without CRs, for a fixed ratio $\phi$, the reduction to the growth rate of the nearly transverse perturbations is more than a case with perturbations along the magnetic field line according to Fig. \ref{fig:f1}. In our analysis, CRs particles are diffusing along the magnetic field lines. Thus, diffusion of CRs is more or less negligible for nearly transverse perturbations which implies a maximum pressure support by CRs in the momentum equation.

The profile of the growth rate of condensation mode reaches to its largest value for a wavenumber $k_{\rm max}$ which depends on the input parameters. Fig. \ref{fig:f1} shows that in the absence of CRs as the direction of the perturbations tends to be along the magnetic field line, the maximum wavenumber $k_{\rm max}$ decreases. But when dynamics of CRs is included, the wavenumber $k_{\rm max}$ depends on the ratio $\phi$ and the  non-dimensional diffusion coefficient $\psi_0$. For nearly transverse perturbations, Fig. \ref{fig:f1} shows that the wavenumber $k_{\rm max}$ decreases when the ratio $\phi$ increases. But this behavior is reversed when perturbations are along the magnetic field lines, i.e. $k_{\rm max}$ increases with the ratio $\phi$.

Fig. \ref{fig:f2} shows the effect of CRs diffusion along the magnetic field lines. All the input parameter are the same as Fig. \ref{fig:f1}, except for $\psi_0$ which varies from $0$ to $100$ and the ratio $\phi$ is kept to be fixed, i.e. $\phi=0.6$. Growth rate of the condensation mode increases with the diffusion of CRs. This destabilizing effect of diffusion is enhanced for the perturbations along the magnetic field lines. For this type of perturbation the maximum wavenumber $k_{\rm max}$ is decreasing as more CRs particles are diffusing. Note that when diffusion of CRs is efficient, the growth rate tends to the profile of a case without CRs. This behavior  is more evident for the perturbations along the magnetic field lines when $\psi_0$ is large according to Fig. \ref{fig:f2}. In other words, compression along the magnetic field lines  due to the magnetothermal instability is stronger when diffusion of CRs are considered.

\section{discussion}
\label{sec:D}

Interstellar CRs can impinge on the structure formation in interstellar medium either by contributing to ionization or constituting as an extra source of heating or even through dynamical coupling to the plasma.  In this study, we analyzed dynamical effects of CRs on the unstable modes in magnetothermal systems. CRs and the plasma are considered as two different fluids and our linear analysis implies a stabilizing effect due to the existence of CRs. Since CRs are diffusing along the magnetic field line, the level of stabilization of the condensation modes decreases, in particular for the perturbations along the magnetic field lines. In fact, in the classical analysis of magnetothermal instability one can show that a purely transverse magnetic field can prevent thermal condensation \citep{field65}. On the other hand, CRs can diffuse along the magnetic field line and not perpendicular to it. Therefore, the magnetic and CRs pressures add up to the thermal pressure for transverse perturbations where diffusion of CRs is negligible. This implies a more magnetothermally stable system. However, CRs pressure reduces for the perturbations along the magnetic field line because of the diffusion of CRs particles. Since the stabilization effect of CRs is stronger for the transverse perturbations, we may expect formation of elongated clouds via thermal instability in the presence of CRs. Effectiveness of CRs on the reduction of the growth rate is anisotropic and CRs diffusion can be a key factor in the final alignment of the elongated clouds.

We can also compare our analysis with similar previous studies like KP83. Although the main goal of KP83 is about possible role of CRs in ISM, in comparison to our study there are significant differences that we summarize them here:

\noindent {(a)} The present paper studies thermal instability, and so, the net {\it cooling function} is considered. But cooling function  is completely neglected in KP83 and the same for the thermal conduction. Adiabatic variations of the gas component is considered in KP83 instead of the complete form of the energy equation (see, eqs. 22 and 23 of KP83). That is a significant difference. Because in our analysis thermal instability is occurring on cooling time scale, but in KP83 there is not thermal effects due to the cooling of the system. So, KP83 can not address {\it thermal condensation modes} with CRs in a typical ISM (such as HI regions) which may have a vital role in theory of star formation. But our analysis discusses about thermal {\it condensation modes} including dynamical role of CRs. Therefore, KP83 and the current paper are addressing somewhat different systems, but leading to a consistent picture for the role of CRs. KP83 showed that CRs operate as an stabilizing factor in Parker instability. Our paper shows that CRs play the same role but within the context of {\it thermal instability} in ISM even at scales comparable to HI regions.

\noindent {(b)} In the light of the above point, our initial state is also different. While we start with a homogenous  initial configuration, KP83 starts with an initial configuration in which the gas is supported in the vertical direction by both {\it gravity} and {\it magnetic} and {\it thermal} pressures. Then, their initial configuration depends on the vertical spatial coordinate (see Figure 1 of KP83). This initial set up is similar to the classical Parker analysis. In our analysis, self-gravity is neglected. Because we are interested in formation of structures due to thermal instability in systems  where self-gravity has a negligible role (like HI regions).

\noindent {(c)} Our initial  state is actually corresponding to a thermal  equilibrium configuration for which the net cooling function is zero. But initial state of KP83 is defined by the balance of different forces in the vertical direction.

\noindent {(d)} Considering the above points, we determined  domains of the stability in Fig. \ref{fig:ff1} numerically (see also equation (\ref{eq:stable})). Since we have considered the complete form of the energy equation including the net cooling function, the stability regions are mainly determined by the net cooling function and the parameters relating to the CRs. We show that when the net cooling of the system is included the stability criteria is not a trivial issue. But since KP83 did not take into account the energy equation for the gas component, their stability criteria has been written in terms of a critical polytropic exponent. Therefore, one can not use their stability condition for a system with a given net cooling function as we have in our study.

We showed that the level of diffusion of CRs particles is determined by the non-dimensional coefficient $\psi_0$. But dependence of this parameter  on the physical parameters of the system is a complicated function of the net cooling function. Having a fixed value for the diffusion coefficient $\kappa_{\parallel}$ around $10^{28}$ cm$^{2}$ s$^{-1}$ \citep{berezinski90, ptuskin2001}, our key parameter $\psi_0$ depends on the physical variables that control the rate of cooling such as temperature and ionization fraction. For simplified cooling systems, we can derive dependence of $\psi_0$ on the parameters of the system. For instance, \citet*{schwarz72} studied formation formation of clouds via thermal instability in ISM. They showed that the cooling time scale of a low-density plasma of cosmic abundance can be approximated as $\tau_{cool}\approx T / nx$ for temperatures within the range of $100$ K to $10^4$ K. Here, $n$ is the number density of particles and $x$ is the fraction density of the electrons, i.e. $x=n_{e}/n$. Then, we have $\psi_{0} \propto \kappa_{\parallel} \rho x /T^2$. Therefore, in cold systems, one may expect an efficient diffusion of CRs along the magnetic field lines which implies negligible dynamical effect of CRs on the condensation mode (see Fig. \ref{fig:f2}). But as the temperature of the system increases, the level of diffusion decreases. Note that $\kappa_{\parallel}$ is assumed to be fixed and independent of the properties of the system in the above argument. However, there are some points that make the problem more complicated. An unstable magnetothermal system may become turbulent at nonlinear regime of the evolution. Then, diffusion of CRs particle may be affected by the level of turbulence within the system.

In the light of our results, we think, the formation of structures in the protogalactic halo environment can be reanalyzed by considering dynamical role of CRs. For example, \citet{baek2006} found that dense clumps first form out of hot background gas by thermal instability in the protogalactic environment. For such systems, the cooling time scale and the sound speed are estimated as $\tau_{cool} = 2\times 10^{7}$ yr and $c_{s}=198$ km s$^{-1}$ \citep{baek2006}. Then, the non-dimensional parameter diffusion parameter becomes $\psi_{0} \approx 0.04$. This implies that the diffusion of CRs along the magnetic field line is going to be negligible and the growth rate of condensation mode is significantly reduced depending on the ratio parameter $\phi$. Therefore, the rate of the structure formation in the protogalactic halo environment via thermal instability is reduced because of the existence of CRs. This encouraging result can be studied further via numerical simulations of thermal instability with CRs for the protogalactic systems.

In another study, \citet*{kimkim2008} studied galactic spiral shock waves with thermal instability. They showed that initially uniform gas rapidly separates into warm and cold phases as a result of thermal instability and also forms a quasi-steady shock that prompts phase transitions \citep{kimkim2008}. Although existence of CRs and their stabilizing effect may not change qualitative scenario of \citet{kimkim2008}, we can expect the initial gaseous disc evolves into warm and cold phases with much slower rate comparing to a similar system  without CRs. This is an interesting research topic for future.

\section*{Acknowledgments}

I thank the anonymous referee for comments leading to a greater clarity of the paper. I am grateful to Luke Drury and Tom Hartquist for their useful comments and suggestions.

\bibliographystyle{mn2e}
\bibliography{myref}

\end{document}